# 4MOST Consortium Survey 7: Wide-Area VISTA Extragalactic Survey (WAVES)

Simon P. Driver[1]
Jochen Liske[2]
Luke J. M. Davies[1]
Aaron S. G. Robotham[1]
Ivan K. Baldry[3]
Michael J. I. Brown[4]
Michelle Cluver[5]
Koen Kuijken[6]
Jon Loveday[7]
Richard McMahon[8]
Martin J. Meyer[1]
Peder Norberg[9]
Matt Owers[10]
Chris Power[1]
Edward N. Taylor[5]
and the WAVES team

[1] International Centre for Radio Astronomy Research/University of Western Australia, Perth, Australia
[2] Hamburger Sternwarte, Universität Hamburg, Germany
[3] Astrophysics Research Institute, Liverpool John Moores University, UK
[4] School of Physics and Astronomy, Monash University, Melbourne, Australia
[5] Centre for Astrophysics and Supercomputing, Swinburne University of Technology, Hawthorn, Australia
[6] Sterrewacht Leiden, Universiteit Leiden, the Netherlands
[7] University of Sussex, Brighton, UK
[8] Institute of Astronomy, University of Cambridge, UK
[9] Department of Physics, Durham University, UK
[10] Department of Physics and Astronomy, Macquarie University, Sydney, Australia

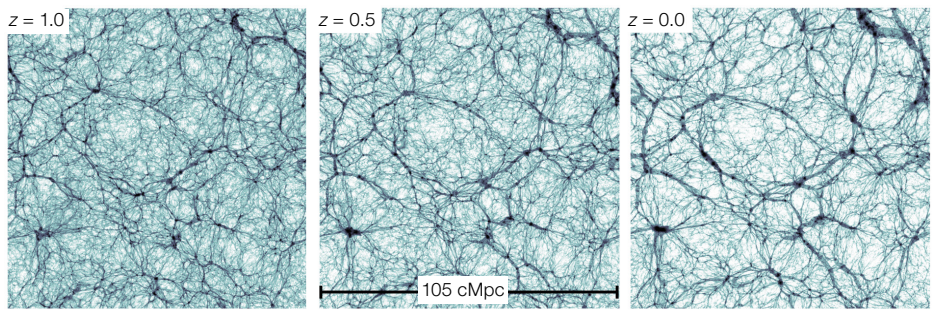

Figure 1. Three panels showing a 105 × 105 comoving Mpc simulated box of the underlying dark matter distribution at three redshifts ($z$). The predicted evolution of the density field results in the movement of mass from voids into filaments, groups and clusters. Measuring this evolution by constructing group, filament and void catalogues is one of the core science goals of WAVES. The figure is by P. Elahi, based on the Synthetic UniveRses For Surveys (SURFS) simulations (Elahi et al., 2018).

WAVES is designed to study the growth of structure, mass and energy on scales of ~ 1 kpc to ~ 10 Mpc over a 7 Gyr timeline. On the largest length scales (1–10 Mpc) WAVES will measure the structures defined by groups, filaments and voids, and their emergence over recent times. Comparisons with bespoke numerical simulations will be used to confirm, refine or refute the Cold Dark Matter paradigm. At intermediate length scales (10 kpc–1 Mpc) WAVES will probe the size and mass distribution of galaxy groups, as well as the galaxy merger rates, in order to directly measure the assembly of dark matter halos and stellar mass. On the smallest length scales (1–10 kpc) WAVES will provide accurate distance and environmental measurements to complement high-resolution space-based imaging to study the mass and size evolution of galaxy bulges, discs and bars. In total, WAVES will provide a panchromatic legacy dataset of ~ 1.6 million galaxies, firmly linking the very low ($z$ < 0.1) and intermediate ($z$ ~ 0.8) redshift Universe.

## Scientific context

The structures which we see in the Universe today, from galaxies to groups, clusters, filaments and voids, were moulded by the underlying dark matter distribution and its hierarchical assembly (Fall & Efstathiou, 1980; Frenk et al., 1988) — without dark matter, large scale structure, galaxies, stars and indeed life would not exist. Numerical simulations of the growth of dark matter structures (for example, Springel et al., 2005) start with initial conditions provided by observations of the cosmic microwave background and then apply only the action of gravity. These simulations successfully explain the very large-scale structure seen in the Universe today, as revealed on Mpc scales and above by surveys such as the Two-degree Field Galaxy Redshift Survey (2dFGRS; Colless et al., 2001), the Sloan Digital Sky Survey (SDSS; York et al., 2000), and the Galaxy And Mass Assembly (GAMA) survey (Driver et al., 2011; Liske et al., 2015).

The latest dark matter simulations (for example, Ishiyama et al., 2015; Ludlow et al., 2016) predict fine details in this structure down to relatively low masses ($10^8 M_\odot$), as well as strong evolution in the growth of large-scale structure on scales of 1 to 100 Mpc over relatively recent times (see Figure 1). This mass flow, from lower to higher density environments, results in the late-time emergence of massive groups and clusters, which can be used to directly trace the evolution in the underlying dark matter distribution (Robotham et al., 2011; Alpaslan et al., 2014).

With most major spectroscopic galaxy surveys (for example, the Dark Energy Spectroscopic Instrument [DESI] survey; DESI Collaboration, 2016) focusing on the sparse sampling of large areas, WAVES[1] is unique in pursuing a fully sampled, high-completeness strategy over relatively modest volumes. This high-density sampling will allow us to directly observe the emergence of fine structure (i.e., groups, filaments and voids), trace the growth of mass and its environmental dependence (i.e., mergers and in-situ star formation; Davies et al., 2015), and follow the primary energy production pathways (i.e., star formation and active galactic nuclei) on kpc to 100 Mpc scales — length scales over which the interaction of dark matter with baryons is strongest.

WAVES will test the Cold Dark Matter (CDM) paradigm by measuring redshifts for ~ 1.6 million galaxies, split between two sub-surveys. WAVES-Wide will cover an area of ~ 1200 square degrees, probing to significantly lower galaxy and halo masses in the low-redshift Universe than previous surveys. WAVES-Deep will cover an area of ~ 70 square degrees to quantify the evolution of structure, mass and energy production over a ~ 7 Gyr baseline. Within the environments probed by WAVES, we will extend studies of the galaxy population and its evolution to lower stellar masses and to more diffuse and lower-surface-brightness systems than was previously possible, while tracing a timespan over which half of the stars in the Universe formed and within which the Hubble sequence emerged. Our goal is to understand the physical processes that drive the evolution of gas to stars and the dependence of these processes on the host galaxy, halo properties and the larger-scale environment.

### Specific scientific goals

Below we briefly list some of the core science goals we plan to pursue with WAVES-Wide and WAVES-Deep. In the following, $Z$ refers to the apparent AB magnitude in the $Z$-band, and $z_{phot}$ to the photometric redshift as estimated from broad-band $ugriZYJHKs$ photometry.

**WAVES-Wide** (~ 0.9 million galaxies with $Z \lesssim 21.1$ magnitudes and $z_{phot} \lesssim 0.2$):
– Identify 50 000 dark matter halos down to a halo mass of ~ $10^{11} M_\odot$ and measure the Halo Mass Function as well as the halo mass–baryonic content relation over 4 orders of magnitude in mass.
– Measure the void distribution function within a representative volume of the Universe (i.e., with sample variance < 5%).
– Measure the length, width and mass content of filaments and tendrils within a representative volume of the Universe (i.e., with sample variance < 5%).
– Quantify the star formation rates, masses and structural properties of central and satellite systems, across a wide range of dark matter halo mass.
– Identify 10 000 Milky-Way-mass ($10^{12} M_\odot$) halos to study galaxy properties in the most typical environment in which most mass resides.

**WAVES-Deep** (~ 0.75 million galaxies with $Z \lesssim 21.25$ magnitudes and $z_{phot} \lesssim 0.8$):
– Identify 20 000 dark matter halos down to a halo mass of ~ $10^{14} M_\odot$ over a broad redshift range and measure the predicted evolution of the high-mass end of the Halo Mass Function.
– Measure the major and minor galaxy merger rates across a broad range of environments and dark matter halo mass to $z \sim 0.8$.
– Quantify the gas, stellar and dust mass growth of galaxies and of their structural components as a function of environment to $z \sim 0.8$.
– Measure the evolution of the cosmic spectral energy distribution, and thus the evolution of energy production in the Universe, to $z \sim 0.8$ by combining WAVES-Deep with complementary X-ray, ultraviolet, optical, infrared and radio imaging data.

### Science requirements

**Completeness:** For both WAVES-Wide and WAVES-Deep we require spectroscopic completeness of > 90% to be able to robustly identify galaxy groups.

**Area:** We require mostly contiguous survey areas of sufficient extent to ensure that sample variance is smaller than 5% at all redshifts (Driver & Robotham, 2010):
– At $z \lesssim 0.2$ we require an area > 1200 square degrees (~ 0.075 Gpc$^3$).
– At $z \sim 0.50$ we require an area > 50 square degrees (~ 0.04 Gpc$^3$ for $\Delta z = 0.05$).
– At $z \sim 0.80$ we require four areas, each with > 4 square degrees (~ 0.015 Gpc$^3$ for $\Delta z = 0.05$).

**Depth:** For WAVES-Deep we aim to probe below the "knee" of the stellar mass function, to ensure we capture the majority of stellar mass at all epochs, out to a redshift of ~ 0.8. This requires a limiting magnitude of $Z \sim 21.25$ magnitudes. For WAVES-WIDE we will go as deep as time allows over the required area (~ 1200 square degrees), resulting in a limit of $Z \sim 21.1$ magnitudes. This provides a 1.3-magnitude improvement over the previous GAMA survey (236 square degrees) and the planned DESI Bright Galaxy Survey (14 000 square degrees).

**Field locations:** WAVES-Wide is confined to the footprint of the Kilo-Degree Survey (KiDS) and the VISTA Kilo-Degree Infrared Galaxy Survey (VIKING), as these surveys provide the data necessary to construct the WAVES-Wide input catalogue. These data are of sufficient depth, resolution and quality to enable robust flux measurements matched to the 4MOST spectroscopic limit, to achieve robust star-galaxy separation, and to provide reliable photometric redshifts to a precision of ± 0.03 (Bilicki et al., 2018), required to select targets with $z_{phot} \lesssim 0.2$.

WAVES-Deep comprises the G23 region of the GAMA survey, because of its extensive multi-wavelength coverage, as well as the central 4 square degrees of each of the four declared Deep Drilling Fields of the Large Synoptic Survey Telescope (LSST). These will be key focus areas for future observations with ground-based radio interferometers and space-based imaging facilities, including the ESA Euclid mission (cf. Figure 2). The exact locations of these deep fields are, however, subject to change.

| Survey region | Right ascension ($\alpha$) (deg) | Declination ($\delta$) (deg) | Area (deg$^2$) | Target selection | Target density (deg$^{-2}$) | No. of targets ($10^3$) |
|---|---|---|---|---|---|---|
| WAVES Wide North (WWN) | $157 < \alpha < 225$ | $-3 < \delta < 4$ | 545 | $Z < 21.1$ | 750 | 410 |
| WAVES Wide South (WWS) | $-30 < \alpha < 52.5$ | $-35.9 < \delta < -27$ | 625 | $z_{phot} \lesssim 0.2$ | 750 | 470 |
| WAVES Deep (WD23) [GAMA23] | $-21 < \alpha < -9$ | $-35 < \delta < -30$ | 50 | $Z < 21.25$ | 11 000 | 550 |
| WAVES Deep (WD01) [ELAIS-S] | 8.95 | –43.70 | 4 | $z_{phot} \lesssim 0.8$ | 11 000 | 45 |
| WAVES Deep (WD02) [XMMLSS] | 35.5 | –5.55 | 4 | | 11 000 | 45 |
| WAVES Deep (WD03) [ECDFS] | 53.0 | –28.0 | 4 | | 11 000 | 45 |
| WAVES Deep (WD10) [E-COSMOS] | 150.12 | 2.50 | 4 | | 11 000 | 45 |
| Total | | | 1236 | | | 1610 |

Table 1. WAVES field locations and areas, selection criteria and input catalogue size.

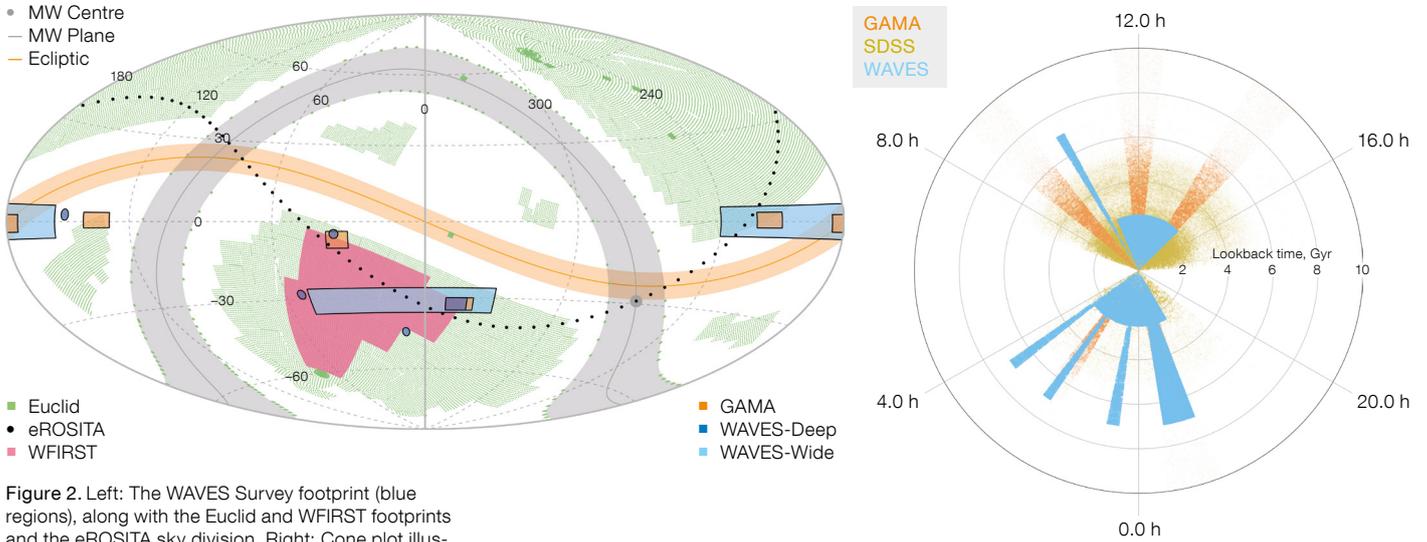

Figure 2. Left: The WAVES Survey footprint (blue regions), along with the Euclid and WFIRST footprints and the eROSITA sky division. Right: Cone plot illustrating the spatial depth of WAVES-Wide, WAVES-Deep and two other prominent spectroscopic galaxy surveys.

### Target selection and survey area

Table 1 specifies our current survey design, indicating the survey regions and the selection criteria we are likely to apply to our input catalogue in terms of the limiting $Z$-band flux ($Z$), and a photometric redshift estimate ($z_{phot}$) based on our $ugriZYJHKs$ data from KiDS and VIKING. An additional criterion is star-galaxy separation, which will likely be based on the $J-Ks$ colour and the measured half-light radius. Objects classified as stars by their colour or size will not be targeted.

The above survey design was derived from a combination of factors which include: detailed simulations of the growth of structure; the need to reduce sample variance to below 5% (Driver & Robotham, 2010); observability with 4MOST; the availability of appropriate data for an input catalogue; and declared future multi-wavelength survey programmes likely to complement and maximise the survey's legacy value (for example, those currently expected from the LSST, the Square Kilometre Array [SKA], Euclid, and the Wide Field Infrared Survey Telescope [WFIRST]).

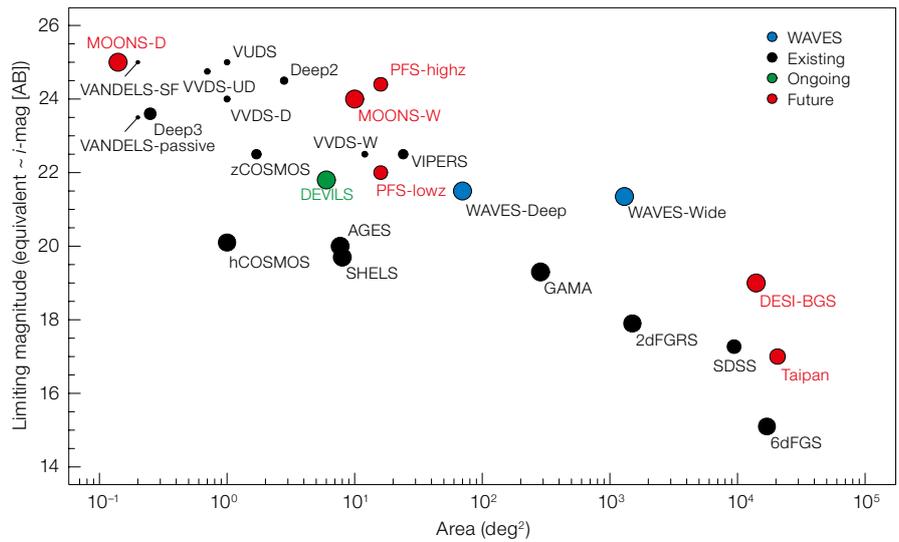

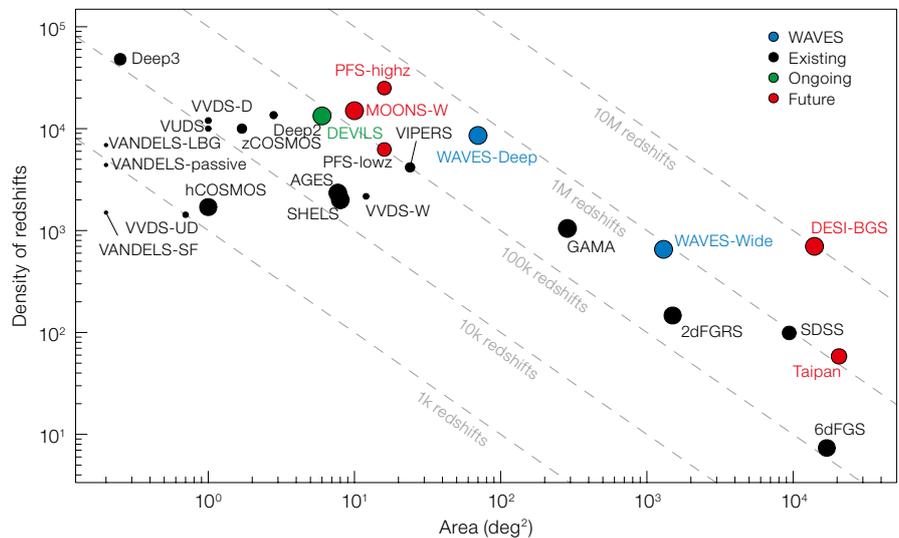

Figure 3. A comparison of recent, ongoing and future surveys showing the competitive edge of WAVES in the parameter space of area, limiting magnitude and target density.

The left panel of Figure 2 shows the WAVES footprint on the sky and its overlap with the forthcoming Euclid and WFIRST space-based imaging programmes. The right panel indicates the improvement of WAVES over the SDSS and GAMA surveys as it pushes to both higher density nearby and to higher lookback times (indicated by the radial axis). Figure 3 shows the competitiveness of WAVES compared to recent, ongoing and future spectroscopic programmes in terms of area, limiting magnitude and target density.

### Spectral success criteria and figure of merit

A WAVES target will be deemed successfully observed if its redshift can be measured with a confidence greater than 90% from its 4MOST spectrum. The ability to construct a high-quality group catalogue, and hence the value of WAVES, depends sensitively on its spectroscopic completeness. We have thus chosen to define the WAVES figure of merit (FoM) purely in terms of completeness and to do so highly non-linearly; the FoM is a power-law function of the completeness, with an exponent of 3 up to a completeness of 0.9, and an exponent of 6 thereafter. The normalisation is chosen such that our requirement of a completeness of 0.9 results in a FoM = 0.5, whereas our goal of a completeness of 0.95 gives a FoM = 1.0.

The above FoM is first defined independently for WAVES-Wide and WAVES-Deep. The overall WAVES FoM is then defined as the lower of the two sub-survey FoMs.

### Acknowledgements

We acknowledge funding from our universities, the Australian Research Council (ARC), the Australian Department of Industry, Innovation and Science (DIIS) and the Deutsche Forschungsgemeinschaft (DFG).

### Links

[1] The WAVES survey: https://wavesurvey.org/

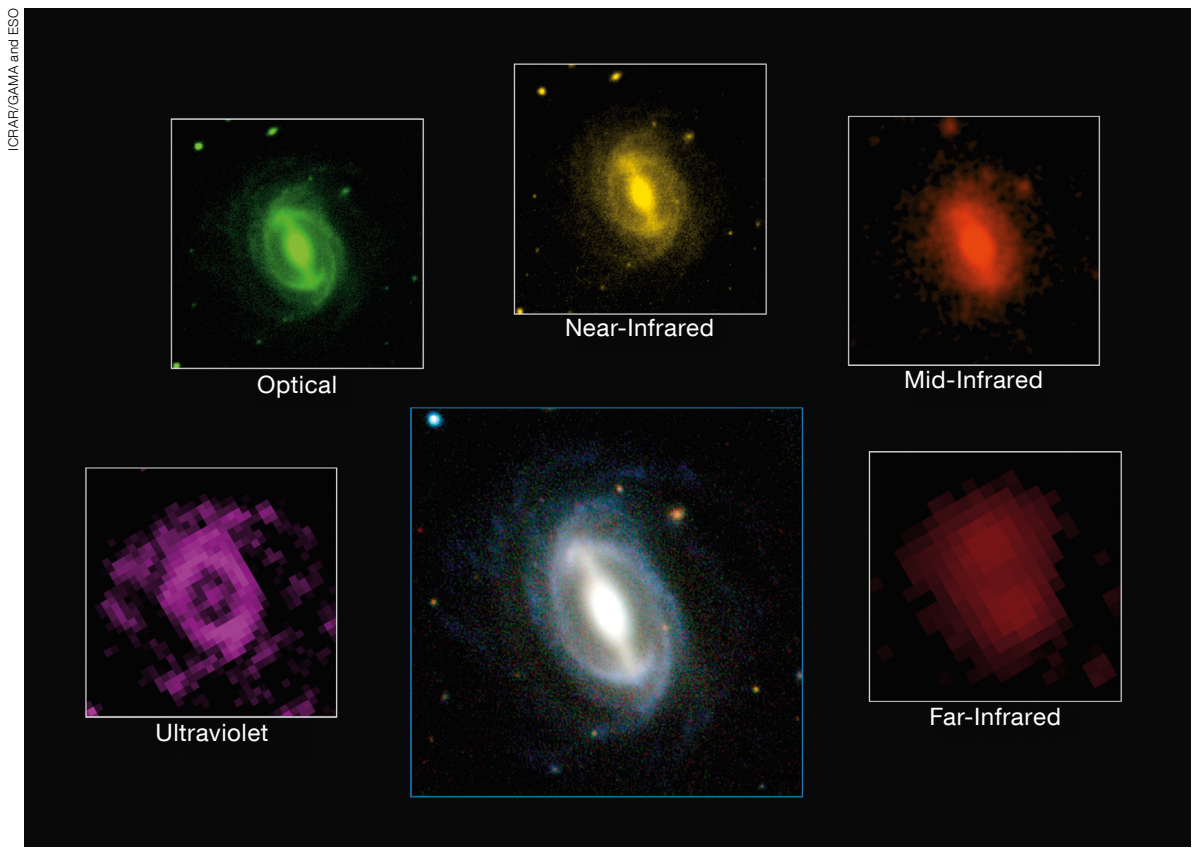

A typical galaxy from the GAMA survey observed at different wavelength regimes.